# UBC/Laval 2.7-METER LIQUID MIRROR TELESCOPE

Paul Hickson[1], Ermanno F. Borra[2], Rémi Cabanac[1,2]

Robert Content[2,3], Brad K. Gibson[1,4], and Gordon A. H. Walker[1]

1. Department of Geophysics and Astronomy, University of British Columbia, 2219 Main Mall, Vancouver, B.C. V6T 1Z4, Canada

2. Département de Physique, Université Laval, Quebec, P.Q. G1K 7P4

3. Present address: Steward Observatory, University of Arizona, Tucson AZ 85721

4. Present address: Department of Astrophysics, University of Oxford, Keble Road, Oxford OX1 3RH, U.K.

ABSTRACT

We report the successful development and operation of a large astronomical liquid-mirror telescope. Employing a rotating 2.7-meter diameter mirror with a surface of liquid mercury, the telescope images a half-degree diameter field centered at the zenith. Located near Vancouver, British Columbia, it is equipped with a low-noise 2048 × 2048-pixel CCD detector, operating in TDI mode, which produces continuous imaging of a 20′ wide strip of sky with 2-minute integration time. Images with FWHM of 2″ or less are regularly obtained. This image quality is limited only by atmospheric seeing and star-trail curvature. The telescope is equipped with a series of narrow-band filters, designed to produce 40-point spectral energy distributions from 0.4 to 1.0 um of all detected objects. These will allow classification and redshift estimation of approximately $10^4$ galaxies and $10^3$ quasars to a limiting magnitude of $R \simeq 21$.

Subject headings: cosmology: observations – surveys – telescopes

# 1. INTRODUCTION

More than a century ago it was realized that the equilibrium configuration of the surface of a uniformly rotating liquid is a paraboloid (Gibson 1991 gives a historical review). If the liquid is mercury, it will focus light and can be used to form an image of distant objects. Over the past ten years, experiments at Laval University, (Borra *et al.* 1992) have resulted in rotating mercury mirrors as large as 1.5-meter diameter capable of producing diffraction-limited images (Borra *et al.* 1989). Key factors in the success of this program were the use of a thin ($\sim 2$ mm) mercury layer, and an air bearing to support the mirror.

As a step toward the development of very large liquid-mirror telescopes, We have designed and constructed an astronomical telescope employing a rotating 2.7-meter diameter liquid mercury mirror. This instrument began astronomical observations in March, 1994, producing images with resolution limited only by atmospheric seeing and latitude dependent effects of star-trail curvature. Although the telescope was built primarily to develop and test technology for large liquid mirrors, it is among the 20 largest optical telescopes in the world, and is capable of producing data of considerable astronomical interest. We have begun a program of narrow-band imaging, designed to search for high-redshift quasars and primeval galaxies. Over the next two years, we plan to obtain spectrophotometry of all objects in a 13 square degree area of sky by means of multi-narrowband observations using 40 filters. The telescope and auxiliary instrumentation is described in §2, and the performance of this instrument is described in §3. In §4 we provide a summary of the scientific program.

## 2. THE TELESCOPE

The most critical component of the telescope is the primary mirror structure, which must stably support a heavy mercury load with minimal flexure and vibration. A major goal of the 2.7-meter project was the development of a mechanical design which might be extended to larger mirrors. Because of the necessity to keep the total weight within manageable limits, particularly as our mirror is supported by an air bearing, high strength-to-weight ratio is of utmost importance. For this reason composite-laminate materials were employed.

Details of the design and construction of the 2.7-meter mirror are described by Hickson, Gibson & Hogg (1993). The mirror consists of a core of aluminum and styrofoam surrounded by a Kevlar skin. Kevlar was chosen because of its high modulus of elasticity and damping coefficient, and relatively low cost. The upper surface of the Kevlar skin is covered by a smooth layer of polyurethane resin formed by spin-casting. The resulting surface, which is parabolic to within a fraction of a mm, is then covered by a mercury layer two millimeters thick. Its surface deflects less than 0.3 mm when loaded with 150 kg of mercury. A thin oxide layer on the surface of the mercury suppresses surface waves and mercury vapor. The resulting mirror has a focal length of 5.00 meters and a focal ratio of 1.89. The entire mirror, including the mercury, weighs 270 kg and has a fundamental resonant frequency of approximately 12 Hz.

The mirror is supported by an air bearing and driven by a thin mylar belt from a synchronous motor, in the same manner as the Laval liquid mirrors. In order to prevent instability (Content 1992, Hickson, Gibson & Hogg 1993), the entire mirror support system must be extremely stiff. Our air bearing is supported by a 10-cm thick 70 kg aluminum disk with three leveling screws resting on steel pads glued to a concrete pier. The screws are driven by a worm-gear reduction system, which allows the bearing to be leveled to within 2 arcsec.

Figure 1 shows a sketch of the telescope, with one tripod leg removed for clarity. A tripod 5.5 meters high supports an instrument-adaptor and laser alignment system. The tripod is made from Aluminum I-beams, with a central brace for increased rigidity. The instrument adaptor incorporates a corrector lens, filter, an electromechanical focussing mechanism, and a $2048 \times 2048$ pixel Loral CCD camera. The corrector lens, designed by C. Morbey of the Dominion Astrophysical Observatory and built by Arnold Optics in Alberta, employs five elements, the largest of which has a diameter of 24 cm. It reduces all optical aberrations, including field distortion, to less than an arcsec over an $0.5°$ field for visible and near-infrared wavelengths. Because the LMT is a zenith-pointing telescope, the CCD is operated in TDI mode (McGraw, Angel & Sargent 1980; Gibson and Hickson 1992), with continuous readout of data. The CCD allows a strip of sky 21 arcmin wide to be scanned, with a sampling interval of 0.62 arcsec/pixel. The integration time is 129 sec – the time taken for an object to drift the length of the CCD. The readout rate is 32,500 pixels per second, which produces over two gigabytes of data per night. The telescope is controlled by a SUN 386i/250 computer which also acquires data from the CCD camera. Images are displayed in real-time and recorded on 8 mm digital video cassettes at the observatory. The data are preprocessed and calibrated at UBC with an HP 9000/730 computer.

The instrument adaptor also incorporates a laser system which facilitates alignment of the corrector and CCD detector. Three small diode lasers are mounted to project parallel beams directly downward onto the mirror. After reflection, these beams converge at the focus, on the optical axis, producing a spot on the CCD. The focus and orientation of the individual lasers beams are adjusted by stopping the rotation of the mirror, and observing the reflection of the beam from a the stationary mercury pool – when the incident and reflected beams are coincident, the beam is exactly vertical. By this technique, the position of the optical axis can be determined with an error of less than 0.3 mm.

The focus mechanism employs a stepping motor, leadscrews, and linear bearings to move the detector, filter, and corrector lens assembly vertically. In order to compensate for thermal expansion of the tripod, temperature sensors attached to the telescope will be monitored by the computer which will provide focus adjustment based on an empirical thermal expansion model. The instrument adaptor accepts 4-in diameter glass filters, which are inserted between the second and third elements of the corrector lens. A rotator stage facilitates accurate alignment of the CCD columns with the direction of sidereal motion.

The telescope is protected from the wind by a cylindrical building, 4.5 meters in diameter and 6.5 meters high. It is made of wood frame construction, with a 15 cm thick reinforced concrete floor. The roof has a 2.9-meter diameter opening, covered by a retractable roof made of interlocking plastic segments. A mezzanine, 3.5 meters high, and a retractable deck provide access to the prime-focus. A plastic spill-container covers the entire observatory floor and extends part way up the walls, to prevent any escape of mercury in case of accidental spillage.

Several factors dictated the choice of observatory site. Because of the experimental nature of this telescope, quick and easy access to the site was deemed to be essential, ruling out a mountaintop location. The telescope must operate in a quiet, vibration-free environment, yet have sufficient support facilities such as power, workshop, security, etc. At the same time it should be as far removed from light pollution as possible, and be located in a temperate climate. A compromise location was chosen, in a rural area 40 km south-east of Vancouver, at an altitude of 120 meters. We are presently considering the option of moving the telescope to the top of a small mountain in the UBC Research Forest, 60 km east of Vancouver. This site, at an altitude of 400 m, offers a darker sky and potentially better seeing.

## 3. PERFORMANCE

Observations commenced in March 1994, following the installation of the wide-field corrector and CCD system. In order to evaluate the performance of the telescope, corrector, CCD, and the atmospheric conditions at the site, a number of trailed star images were obtained. These were made by scanning the CCD at a rate slightly faster than the sidereal rate, causing the images to be trailed by about $45''$. An example is shown in Figure 2. The effect of atmospheric seeing is to blur the trailed images. The seeing can be estimated from the width of the image in the N-S direction. This width was almost always less than $2''$ FWHM for observations taken over a period of several months.

Aberrations due to the telescope, such as misalignment of the mirror or corrector, appear as a field-angle dependent smearing of the images. Such aberrations are well below $1''$, and are thus negligible compared to the seeing. A second effect that can occur with liquid-mirrors is periodic motion of the star image, synchronous with the rotation of the primary mirror. Such motion can occur if the mirror is not properly balanced, and manifests itself as a small periodic wiggle of the star trails. Our mirror was balanced by adding a small counterweight to the rim and noting the effect on the star trails. In this manner it was possible to reduce the periodic error to a fraction of an arcsec, to the point where it can no longer be detected and has no measurable effect on image quality.

The final component of image distortion results from the TDI operation. This has two main components. The first is due to curvature of the star trails, the second is due to the change in sidereal tracking rate with declination. These errors are latitude-dependent, and increase in proportion to the square of the angular field of view. For our system, they result in image smear of 1-2″, depending on the declination of the object. Star-trail curvature can be distinguished from other types of problems, such as distortion caused by misalignment of the corrector, in that the trails are all slightly concave, in the north direction, by an amount which is essentially constant over the entire field.

Figure 3 shows a typical 2048 × 2048-pixel TDI image in a 33.3 nm wide bandpass centered at 702.4 nm. The faintest objects in this image have $R \simeq 21$. In Figure 4, we have plotted a typical stellar intensity profile. The smooth line is a Gaussian profile with the same FWHM. It is apparent that the image has a sharp near-gaussian core, but has a significant excess flux at radii greater than a few arcsec. This halo is stronger than expected for either a conventional mirror, or a liquid mirror (Borra *et al.* 1992). Further investigation is underway to determine the cause, and to attempt to reduce, the strength of the halo.

## 4. SURVEY PROGRAM

Many questions of fundamental importance, such as the formation and evolution of galaxies and quasars, and the large scale geometry and structure of the universe, can be addressed by surveys of distant objects. Because of their potential for providing very large apertures at relatively low cost, and because they operate in a scanning mode (they cannot be pointed at individual objects), LMTs promise to be ideal instruments to provide the kind of data that are required.

The primary scientific program of the 2.7-m LMT is to study distant galaxies and quasars by means of narrow-band imaging. We have 40 narrowband filters, each having a bandwidth of 4.59% (0.02 dex), covering the entire detectable spectrum from 0.4 - 1.0 um wavelength with uniform logarithmic sampling. Observations are made with a single filter each night. After 40 nights of observing, one obtains a spectral energy distribution for each object. Photometric calibration will be derived from a sequence of secondary spectrophotometric standards in this region of sky. These data will be used to classify objects, and to estimate their redshifts. Analysis of simulated data indicates that, for a signal-to-noise ratio $S/N \geq 5$, redshifts of normal galaxies can generally be determined from such data, with a typical accuracy of $< 3\%$ (Hickson, Gibson and Callaghan 1994, Cabanac 1992).

With a typical image size of 2″ FWHM, we expect that the 2.7m LMT survey will detect galaxies, with $S/N \geq 3$, to a limit of about 20 in integrated galaxy V-magnitude. Over the 13 square degrees for which we expect to achieve complete spectral coverage, there are about 25,000 galaxies with this magnitude or brighter. In the winter observing season, we expect to observe all of these galaxies through about 20 narrowband filters, over about a three month period, based on weather statistics (Hickson and Bennett 1989). In two years, we should obtain complete coverage of the optical band at a resolution $\lambda/\Delta\lambda \sim 22$, giving us spectrophotometry for a large number of galaxies.

The bulk of galaxies with $V \leq 20$ are expected to have have redshifts $z \leq 0.5$.

This is corresponds to a light-travel time of up to ∼ 45% of the age of the universe. This is a sufficiently long time for normal stellar evolution to significantly alter the spectral energy distributions of many galaxies (Rocca-Volmerange & Guiderdoni 1988). In addition, dynamical interactions can induce bursts of star formation, modifying galaxy spectra. There is ample evidence that such evolution is occurring in many (but not all) galaxies even at these modest redshifts (Carlberg 1992; Lavery, Pierce & McClure 1992). The primary goals of the galaxy survey program are to estimate the luminosity function of various galaxy types, to a redshift of $z \simeq 0.3$, and to study the distribution of galaxies on very large ($\gtrsim$100 Mpc) scales.

The survey should also detect many quasars. At a magnitude of $V = 21$, the surface density of quasars is about 100 per square degree. Thus we expect to have about 1000 quasars in the 13 square degree survey area. This is a much larger sample than previous similar surveys (Hartwick & Schade 1990). The narrow band filter data will allow the selection of quasar candidates which have strong emission lines or colors different from galactic stars. Previous studies indicate that redshifts accurate to $\Delta z = \pm 0.02$ should be possible (Drinkwater 1988). The primary objectives of the quasar survey program are to measure the degree of clustering of high redshift (z ∼ 2-3) quasars, and to study the large-scale homogeneity of the quasar distribution.

The narrowband survey will be well suited to a search for primeval galaxies (De Propris *et al.* 1993). We will be sensitive to Lyman-α for a redshift range $2.5 < z < 6.5$. Our estimated 3σ Lyman-α flux limit, for a single scan, is $2 \times 10^{-16}$ erg s$^{-1}$ cm$^{-2}$. This, in combination with the very large survey area and wavelength coverage, will probe the number-flux diagram for faint emission line objects to new limits. We now have in hand scans of approximately 20 square degrees of sky at high galactic latitude, at 11 different wavelengths. These data, which are now being reduced, should provide the first scientific results from the telescope later this year.

This work was supported by grants from the Natural Sciences and Engineering Research Council of Canada. Borra, Cabanac and Content also acknowledge support from the FCAR of the Province of Quebec.

TABLE 1
Telescope Specifications

| | |
|---|---:|
| mirror diameter (cm): | 265 |
| effective mirror area (cm2): | 51755 |
| focal length (cm): | 500 |
| focal ratio: | 1.89 |
| image scale (arcsec/mm): | 41.25 |
| pixel size (microns): | $15 \times 15$ |
| pixel size (arcsec): | 0.619 |
| detector width (NS, pixels): | 2048 |
| detector width (arcmin): | 21.1 |
| detector height (EW, pixels): | 2048 |
| observatory latitude: | 49° 03′ 36″ |
| scan rate (arcsec/sec): | 9.83 |
| readout rate (pixels/sec): | 32471 |
| integration time (sec): | 128.9 |
| filter transmission (700 nm): | 0.95 |
| filter bandwidth (%): | 4.2 |
| sky brightness (R mag/sq arcsec): | 19.2 |
| seeing (typical FWHM, arcsec): | 2.0 |
| available sky area (sq deg): | 83.0 |
| sky area per 8 hr scan (sq deg): | 27.7 |
| area completely covered in 3 months (sq deg): | 13.0 |

Figure Captions

Figure 1. Sketch showing the primary components of the 2.7-meter liquid-mirror telescope.

Figure 2. Trailed star images. These were obtained by using the CCD in TDI mode, with a rate that differed from the sidereal rate. The width of the trails is typically less than 2" FWHM. There is no visible image motion other than that induced by the atmosphere. The trails are curved by an amount corresponding to a maximum imaged displacement of 1" over the extent of our detector, as a result of the northerly latitude of the observatory.

Figure 3. Narrowband TDI image of a field at $15^h\ 29^m\ +49°\ 14'$ (1950) obtained with a single scan by the telescope. The photograph has dimensions of approximately $19'' \times 19''$, and has North up and East to the right. The bright star is SAO 045572. The effective integration time is 129 s. The bandwidth is 33.3 nm with a central wavelength of 702.4 nm. Stars and galaxies can be detected in the digital image to a limiting magnitude of $R \simeq 21$.

Figure 4. Intensity profile of a typical star image. The solid line is a gaussian fit to the central part of the profile.

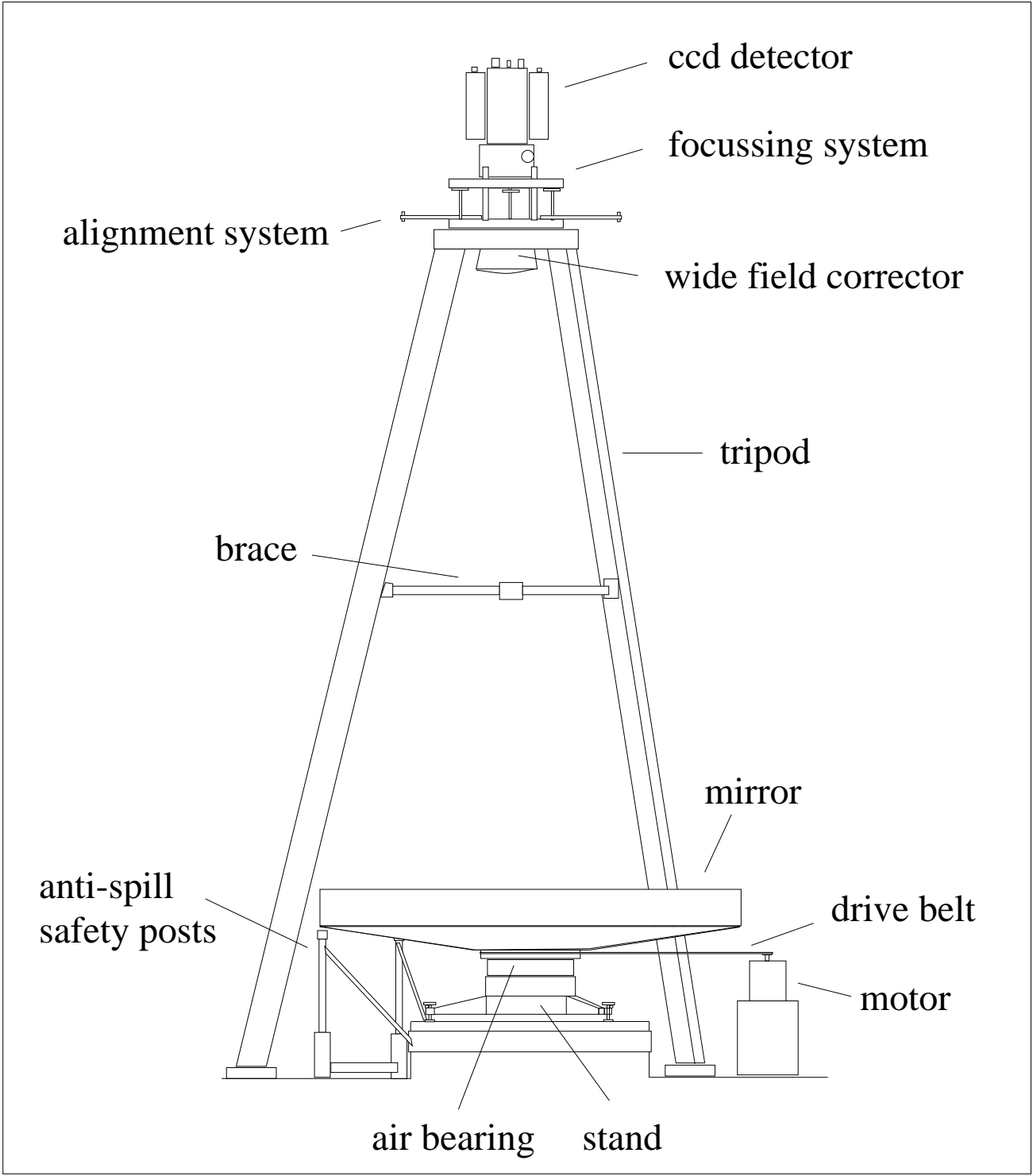

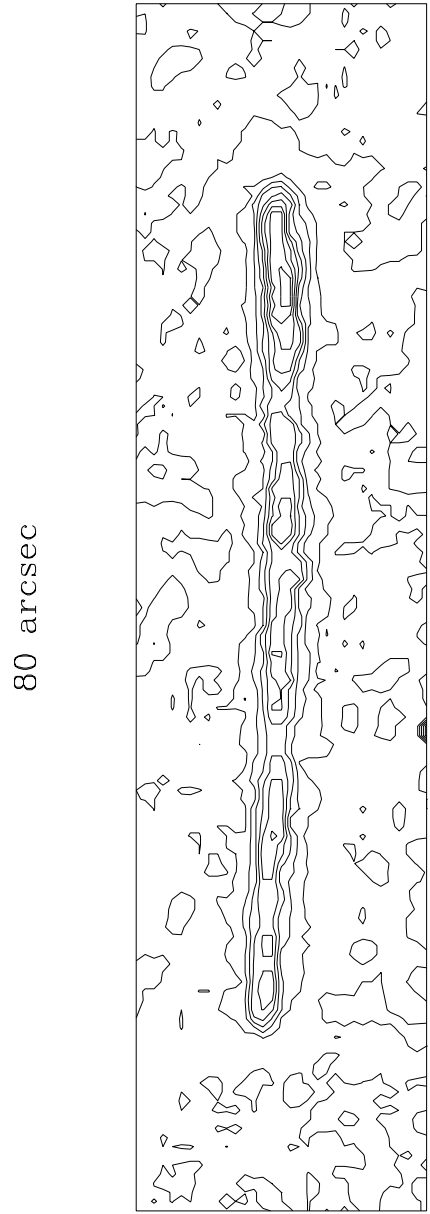

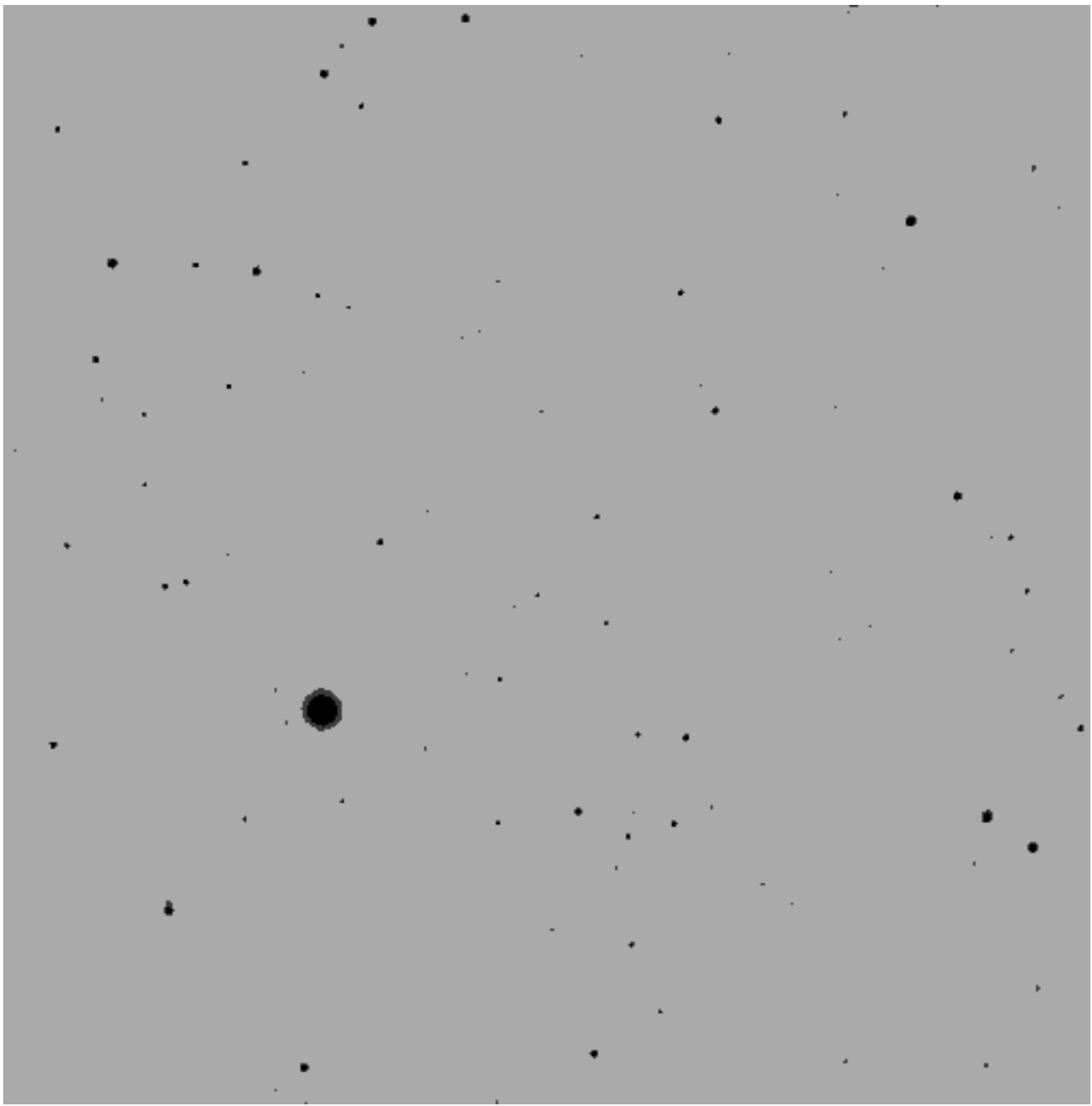

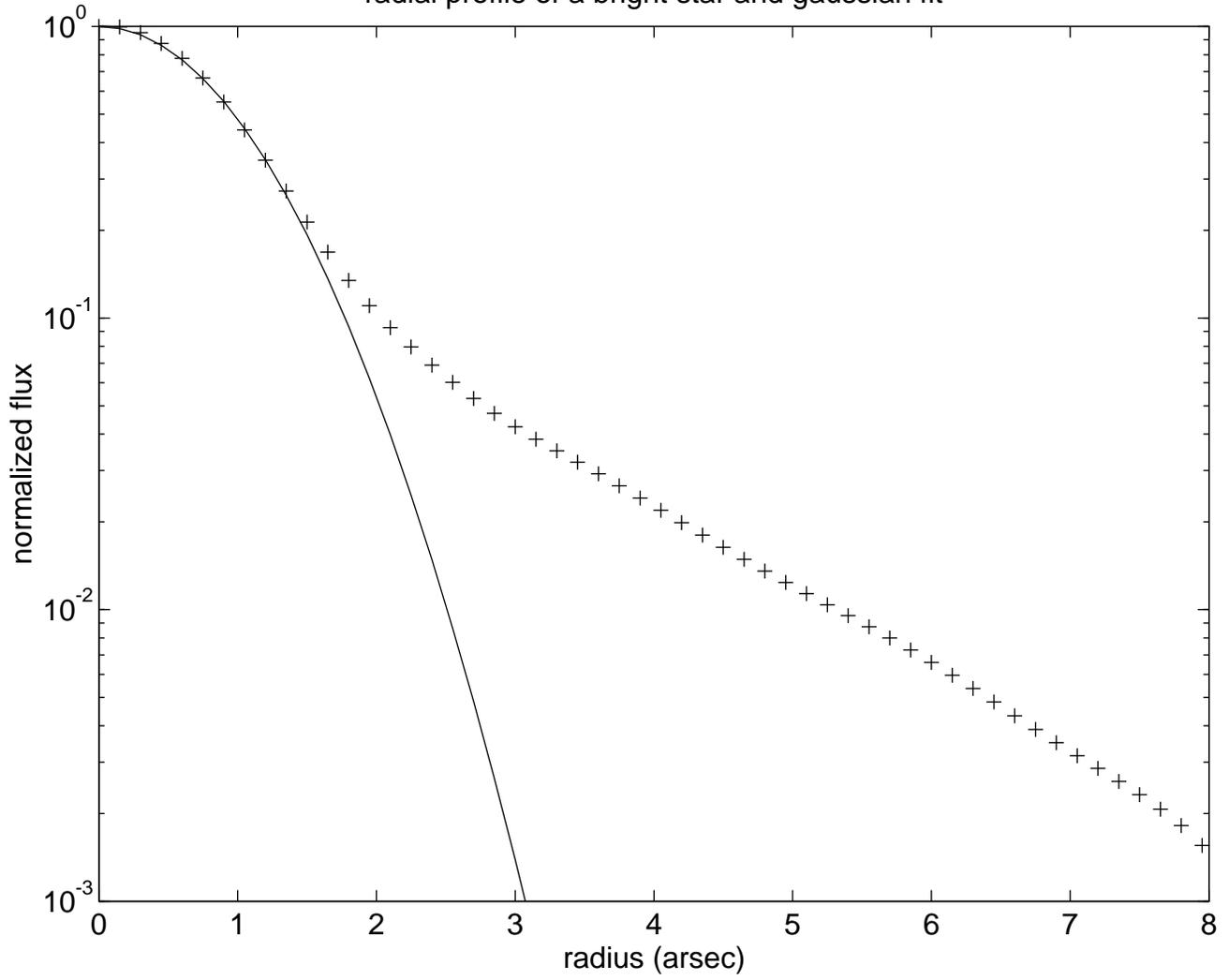

radial profile of a bright star and gaussian fit